\documentclass[twocolumn,notitlepage,showkeys,showpacs,longbibliography]{revtex4-1}
\usepackage[utf8]{inputenc}
\usepackage{stmaryrd}
\usepackage{amsmath}
\usepackage{amssymb}
\usepackage{graphicx}
\usepackage{dcolumn}
\usepackage{bm}
\usepackage{multirow}
\usepackage[colorlinks,linkcolor=blue,citecolor=blue,urlcolor=blue]{hyperref}
\usepackage{color}
\usepackage{ulem}
\usepackage{comment}
\usepackage{gensymb}

\begin{document}
\title{Momentum-inversion symmetry breaking on the Fermi surface of magnetic topological insulators}
\author{Hengxin Tan}
\author{Daniel Kaplan}
\author{Binghai Yan}
\affiliation{Department of Condensed Matter Physics, Weizmann Institute of Science, Rehovot 7610001, Israel}
\date{\today}

\begin{abstract}
Magnetic topological insulators (MnBi$_2$Te$_4$)(Bi$_2$Te$_3$)$_n$ were anticipated to exhibit magnetic energy gaps while recent spectroscopic studies did not observe them. Thus, magnetism on the surface is under debate. In this work, we propose another symmetry criterion to probe the surface magnetism. Because of both time-reversal symmetry-breaking and inversion symmetry-breaking, we demonstrate that the surface band structure violates momentum-inversion symmetry and leads to a three-fold rather than six-fold rotational symmetry on the Fermi surface if corresponding surface states couple strongly to the surface magnetism. Such a momentum-inversion symmetry violation is significant along the $\Gamma-K$ direction for surface bands on the (0001) plane.  
 
\end{abstract}
\maketitle

\section{Introduction}

The intrinsic magnetic topological insulator MnBi$_2$Te$_4$ and its sister compounds (MnBi$_2$Te$_4$)(Bi$_2$Te$_3$)$_n$ ($n = 0,1, 2, 3$) \cite{gong2019experimental,otrokov2019prediction,li2019intrinsic,PhysRevLett.122.206401} bring great opportunities for the realization of quantum anomalous Hall effect (QAHE) \cite{chang2013,deng2020science,nwaa089,deng2021high} and axion insulator (AI) \cite{sciadv.aao1669,PhysRevLett.120.056801,liu2020robust}.
Despite a magnetic surface gap being predicted in theory \cite{PhysRevLett.122.107202,wu2019natural,PhysRevB.102.035144,PhysRevB.102.245136,zhong2021light,PhysRevX.10.031013}, most angle-resolved photoemission spectroscopy (ARPES) experiments, however, observed a gapless surface \cite{PhysRevX.9.041038,PhysRevX.9.041039,PhysRevX.9.041040,PhysRevX.9.041065,PhysRevB.101.161109,XU20202086,PhysRevX.10.031013,hu2020van,PhysRevB.101.161113,PhysRevB.102.045130,PhysRevLett.126.176403}.

To address the discrepancy between experiment and theory, several candidate mechanisms based on either surface magnetic reconstruction, surface structural relaxation, or surface-bulk band hybridization have been proposed \cite{PhysRevX.9.041038,PhysRevB.101.161109,PhysRevX.9.041040,PhysRevX.11.021033,PhysRevB.102.245136,acs.nanolett.0c00031,acsnano.0c03149,shikin2020nature,shikin2021sample,yan2021elusive,garnica2022native}, among which the surface magnetic reconstruction is most extensively discussed. The ground state of bulk materials has the A-type anti-ferromagnetic configuration [intralayer ferromagnetic (FM) and interlayer anti-ferromagnetic (AFM)] for $n=0,1,2$ and FM configuration for $n\ge3$ due to the intensely weakening of the interlayer AFM super-superexchange coupling \cite{klimovskikh2020tunable,hu2020sciadv,wu2020toward}. The surface magnetism was anticipated to open an energy gap on the surface Dirac cone.
Because most ARPES experiments observed no surface gap, it is speculated that the surface magnetic order may be changed to disorder, in-plane AFM (the magnetic moment is along the in-plane direction), or G-type AFM ($i.e.$ AFM along all lattice vectors) (see, for example, Refs.  \onlinecite{PhysRevX.9.041038,PhysRevX.10.031013,PhysRevB.101.161109}). However, recent spectroscopy experiments \cite{PhysRevLett.125.037201,PhysRevLett.125.117205}  claimed that the bulk A-type AFM is robust on the surface. Thus, surface magnetism is still debated.

In this work, we propose to probe the surface magnetism via the symmetry of surface states besides searching for the magnetic gap.
Because the magnetic surface naturally breaks the inversion symmetry and time-reversal symmetry ($\mathcal{T}$), the surface band structure ubiquitously violates the momentum-inversion symmetry, for example, in the chiral surface Fermi arcs of a magnetic Weyl semimetal \cite{liu2018giant,liu2019magnetic,science.aav2334}. 
On the (0001) surface of these magnetic materials, we find that the Fermi surface exhibits a three-fold rotational symmetry if it couples strongly with the out-of-plane magnetism. Otherwise, the Fermi surface shows a six-fold rotational symmetry. 
In the surface Brillouin zone (BZ), such momentum-inversion symmetry breaking is significant along the $\Gamma-K$ direction but vanishes along $\Gamma-M$ due to symmetry constraint. We take MnBi$_4$Te$_7$ ($n=1$) as an example to show the Fermi surface symmetry breaking on both MnBi$_2$Te$_4$- and Bi$_2$Te$_3$-terminated surfaces. 


\section{method \label{method}}
We performed density-functional theory (DFT) calculations with the plane-wave basis set, as implemented in Vienna $ab$-$initio$ Simulation Package (VASP) \cite{VASP,PRB54p11169}. The cutoff energy for the plane-wave basis set is 350 eV.
The generalized gradient approximation parameterized by Perdew-Burke-Ernzerhof \cite{PRL77p3865} is used as the exchange-correlational functional between electrons. A Hubbard $U$ of 5 eV is used for Mn-$d$ electrons.
The bulk BZ is sampled by a $k$-grid of 12$\times$12$\times$4. The zero damping DFT-D3 van der Waals correction \cite{DFTD3} is included in all calculations. Spin-orbit coupling is considered in electronic structure calculations.
The Wannier Hamiltonian is obtained by the maximally localized Wannier functions method \cite{PhysRevB.56.12847} as implemented in \textsc{wannier90} package \cite{wannier90} which is interfaced to VASP. 
The constant-energy contours (Fermi surfaces) of the surface states of MnBi$_4$Te$_7$ (A-type AFM configuration) are calculated by a tight-binding scheme with thick slab models (more than 20 nm) \cite{Tan2022mbt} constructed from the bulk Wannier Hamiltonian.
The surface states are identified by projecting wave functions to the top three surface van der Waals layers.

\section{Results and discussion}
\subsection{Symmetry analysis of surface states \label{sym}}
\begin{figure}[tbp]
\includegraphics[width=0.9\columnwidth]{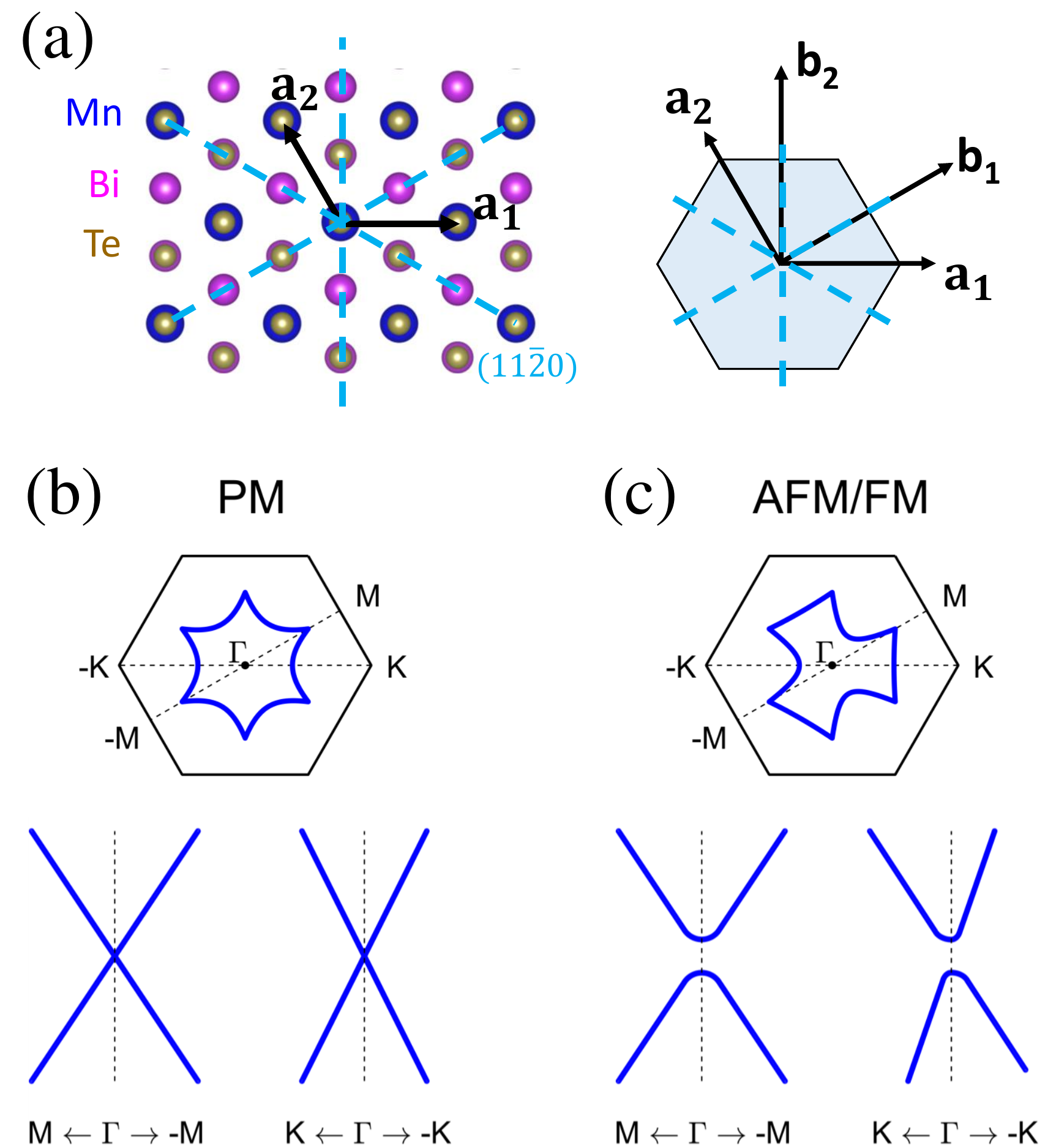}
\caption{\label{schematics} Symmetry, schematic Fermi surface and surface states of the (0001) surface. (a) shows the top view of the MnBi$_2$Te$_4$ surface structure (left panel) and the direction relationship between the lattice vectors $\mathbf{a_1/a_2}$, reciprocal lattice vectors $\mathbf{b_1/b_2}$ and mirror planes in the first Brillouin zone (right panel). The cyan dashed lines represent three mirror planes projected onto the surface plane which are correlated by the three-fold rotational operation [the (11$\bar{2}$0) plane is labeled]. (b) shows the Fermi surfaces (upper) and surface states (lower) of the (0001) surface under the paramagnetic (PM) configuration, which have an equivalent $\mathcal{C}_{6v}$ symmetry. (c) is similar to (b) but for A-type antiferromagnetic/ferromagnetic (AFM/FM) case, where the Fermi surface has only equivalent $\mathcal{C}_{3v}$ symmetry with equivalent $M$ and $-M$ but non-equivalent $K$ and $-K$. Notice that A-type AFM/FM also opens a band gap for the surface states in (c).
}
\end{figure}

Let us start from the symmetry of a perfect (0001) surface without considering the magnetism (i.e., paramagnetic case) in the first stage.
The perfect (0001) surface has $\mathcal{C}_{3v}$ symmetry which is generated by a three-fold rotational symmetry $\mathcal{C}_{3}$ around the out-of-plane axis and a mirror symmetry $\mathcal{M}$ with respect to, e.g. the (11$\bar{2}$0) crystallographic plane as shown in Fig. \ref{schematics}(a). In the momentum space, the $\Gamma-M$ line aligns inside the mirror plane. 
If $\mathcal{T}$ is maintained, the band energy $\varepsilon_n(\bm{k})$ is symmetric to $\varepsilon_n(-\bm{k})$. 
Combining $\mathcal{C}_3$ and $\mathcal{M}$, the band dispersion and the Fermi surface exhibit $\mathcal{C}_{6v}$ symmetry, as manifested schematically in Fig. \ref{schematics}(b).

Next we consider the A-type AFM where $\mathcal{C}_3$ holds. Time-reversal symmetry breaking eliminates the equivalence of $\varepsilon_n(\bm{k})$  and $\varepsilon_n(-\bm{k})$ at generic $\bm{k}$ (e.g., along the $\Gamma - K$ line). 
Thus, the $\mathcal{C}_{6v}$ symmetry of the Fermi surface is lifted.
We note that a combined symmetry by $\mathcal{M}$ and $\mathcal{T}$ still preserves $\varepsilon_n(\bm{k}) =\varepsilon_n(-\bm{k})$ along the $\Gamma-M$ line. 
Therefore, the band structure is symmetric along $\Gamma - M$ direction while it is asymmetric along $\Gamma - K$ direction, as illustrated in Fig. \ref{schematics}(c). 
The combined $\mathcal{MT}$ symmetry is equivalent to a new mirror reflection with respect to the 
$\Gamma - K$ line in the Fermi surface. Together with $\mathcal{C}_3$, the Fermi surface exhibits $\mathcal{C}_{3v}$ symmetry. 

\begin{figure*}[tbp]
\includegraphics[width=1.95\columnwidth]{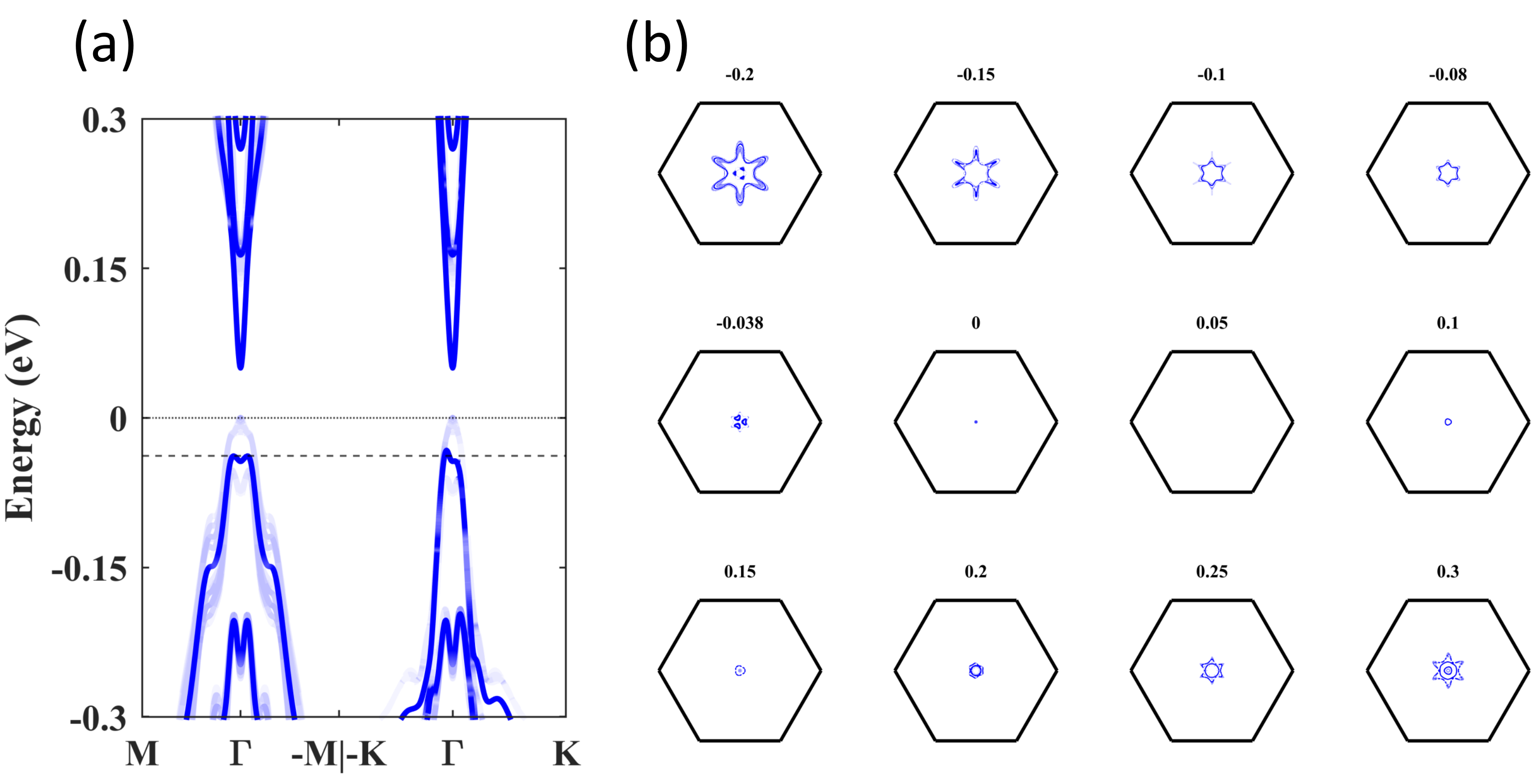}
\caption{\label{MBT147_SL} Surface states and constant-energy contours of the MnBi$_2$Te$_4$-terminated surface of MnBi$_4$Te$_7$. (a) shows the surface states along $\Gamma-M$ and $\Gamma-K$ directions. The color blue stands for the weight of the surface (the larger the weight the darker the color). The surface states along $\Gamma-M$ line are symmetric showing the ``mirror symmetry'' $\mathcal{MT}$ while no such symmetry is shown along $\Gamma-K$ line.
(b) shows the constant-energy contours of the surface states in the full Brillouin zone (black hexagon) at different energies. The title of each panel shows the corresponding energy in eV.
The surface states and constant-energy contours show that, while the $\mathcal{C}_{6v}$ symmetry is most obviously broken near the top of the surface valence bands ($-0.038$ eV) as indicated by the dashed line in (a), the $\mathcal{C}_{6v}$ symmetry is well maintained at other energies $e.g.$, near the experiment Fermi level ($\sim 0.3$ eV). Notice that the maximum of the bulk valence bands is set to energy zero.
}
\end{figure*}

As we mentioned above, some other magnetic configurations were also proposed, such as the in-plane AFM and G-type AFM \cite{PhysRevX.9.041038,PhysRevX.10.031013,PhysRevB.101.161109}. For the same reason of $\mathcal{T}$-breaking, these cases also violate momentum-inversion symmetry on the Fermi surface, where $\mathcal{C}_3$ is further broken. 

Besides the magnetic gap, we can monitor the surface magnetism by the $\mathcal{C}_{6v}$ symmetry-breaking of the Fermi surface. 
The magnitude of symmetry breaking indicates the coupling strength between topological surface states and surface magnetism.  We stress that the momentum-inversion breaking is a general symmetry criterion and applies to general magnetic surface states. For example, the Fermi arcs reported in the recent magnetic Weyl semimetal Co$_3$Sn$_2$S$_2$ \cite{liu2018giant,liu2019magnetic,science.aav2334}) break the momentum-inversion symmetry \cite{PhysRevB.97.235416}. However, such symmetry-breaking was rarely appreciated in recent studies on (MnBi$_2$Te$_4$)(Bi$_2$Te$_3$)$_n$.

\subsection{Surface band structures \label{slab}}

\begin{figure*}[tbp]
\includegraphics[width=1.95\columnwidth]{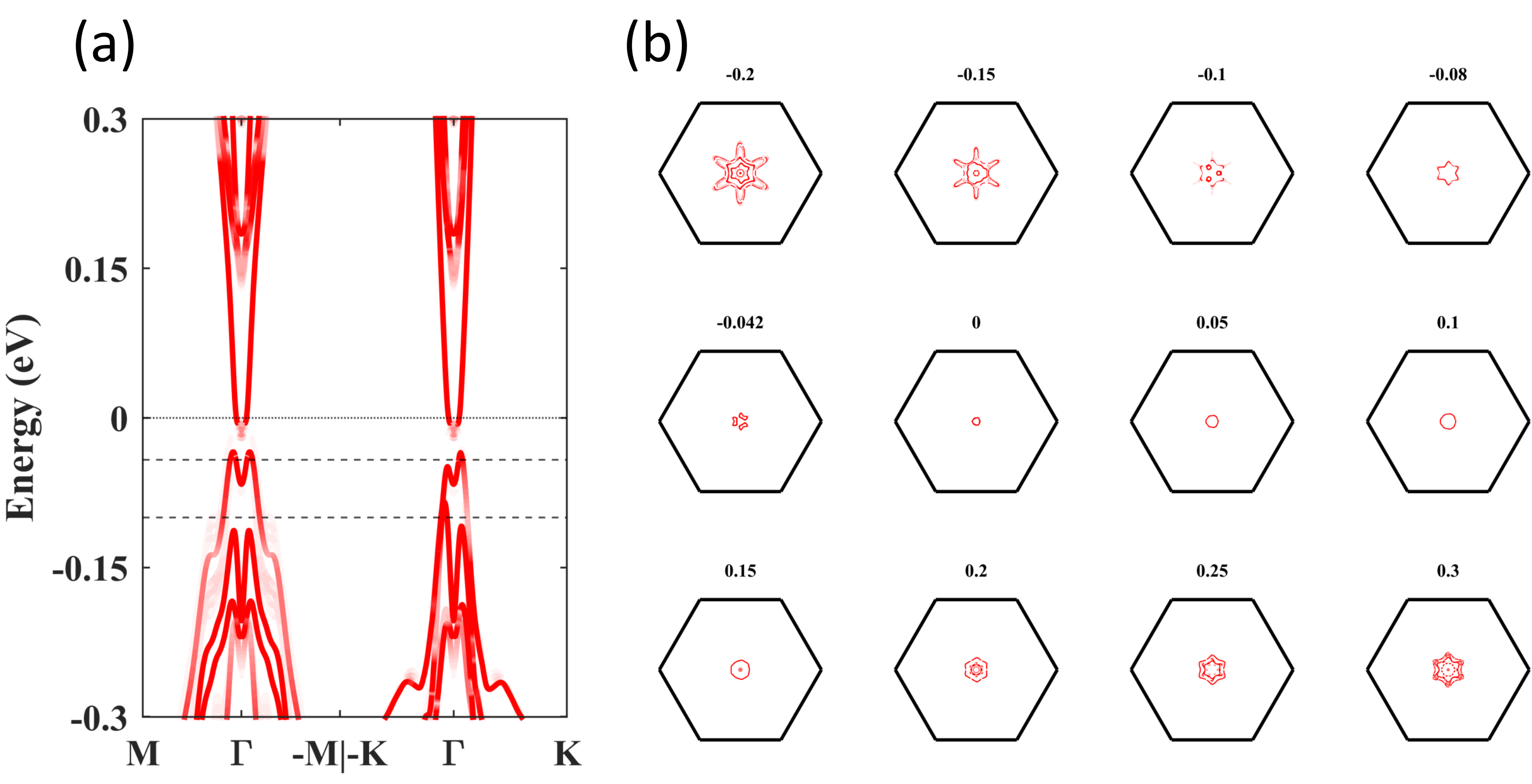}
\caption{\label{MBT147_QL} Similar to Fig. \ref{MBT147_SL} but for the Bi$_2$Te$_3$-terminated surface of MnBi$_4$Te$_7$: (a) surface states in red (the larger the weight of the surface the darker the red color) and (b) constant-energy contours at different energies. Similar to the MnBi$_2$Te$_4$ surface in Fig. \ref{MBT147_SL}, the surface states and constant-energy contours indicate only $\mathcal{C}_{3v}$ symmetry of the surface states while the $\mathcal{C}_{6v}$ symmetry is well maintained at many energies, $e.g$. the experiment Fermi level. The most prominent symmetry breaking happens not only at the top of the surface valence bands ($-0.042$ eV) but also at the energy of about $-0.1$ eV, as indicated by the dashed lines in (a). The maximum of the bulk valence bands is set to energy zero.
}
\end{figure*}

Taking MnBi$_4$Te$_7$ as an example, we now demonstrate our symmetry analysis above by first-principles calculations with thick slab models under A-type AFM.
The top and bottom surfaces of the slab are terminated by MnBi$_2$Te$_4$ and Bi$_2$Te$_3$ van der Waals layers, respectively. The interaction between top and bottom surfaces is avoided in the thick slab \cite{Tan2022mbt}.


\textit{MnBi$_2$Te$_4$-terminated surface.}
The surface states of the MnBi$_2$Te$_4$ surface of MnBi$_4$Te$_7$ are shown in Fig. \ref{MBT147_SL}(a). The surface states near the Fermi energy open a magnetic gap (in the range of about $50 \sim 60$ meV) which comes from the surface Dirac cone (located at the bulk band gap) due to the $\mathcal{T}$-symmetry breaking on the (0001) surface, consistent with previous theoretical calculations.

Figure \ref{MBT147_SL}(a) shows that the surface states along $\Gamma-M$ line are symmetric while however, the surface states along the $\Gamma-K$ line are asymmetric, consistent with our symmetry analysis in Sec.\ref{sym}.
We notice that the most prominent symmetry breaking happens on the top of the surface valence bands (in the energy range of $-43 \sim -33$ meV relative to the maximum of the bulk valence bands), while the symmetry breaking at other energies is less apparent.
In the energy range of $-43 \sim -33$ meV, the surface valence bands are mainly contributed by the topmost MnBi$_2$Te$_4$ layer and therefore couple strongly with magnetism. 
Fermi surfaces in Fig. \ref{MBT147_SL}(b) show more clearly the symmetry of surface states. For example, the constant-energy contour at $-40$ meV shows clearly $\mathcal{C}_{3v}$ symmetry.
Away from the top of the surface valence bands, Fermi surfaces display nearly $\mathcal{C}_{6v}$ symmetry because corresponding states are mainly contributed by the underlying Bi$_2$Te$_3$ layer. 
 For example, the $\mathcal{C}_{6v}$ is well preserved at $\sim 0.3$ eV,

\textit{Bi$_2$Te$_3$-terminated surface.} Similar symmetry breaking behaviors are also confirmed for the Bi$_2$Te$_3$ surface as shown in Fig. \ref{MBT147_QL}. The most obvious symmetry breaking happens not only at the top of the surface valence bands (in the energy range of $-49 \sim -35$ meV, relative to the bulk valence band maximum), but also at the energy range of $-109 \sim -84$ meV, as indicated in Fig. \ref{MBT147_QL}(a).
While the symmetry breaking surface state near $-100$ meV is equally contributed by both the topmost Bi$_2$Te$_3$ layer and MnBi$_2$Te$_4$ layer beneath, the symmetry breaking branch near $-40$ meV is solely contributed by the MnBi$_2$Te$_4$ layer below the surface Bi$_2$Te$_3$ layer.
This is not strange because the magnetism comes from the Mn sublayer in the MnBi$_2$Te$_4$ layer.
The symmetry breaking of the surface band structures near these energies is more clearly shown in Fig. \ref{MBT147_QL}(b).
In addition, the $\mathcal{C}_{6v}$ symmetry at other energies is well maintained.
We note that the surface gap is filled with the bulk valence bands (filtered by the surface projection in Fig. \ref{MBT147_QL}) as discussed in the previous work \cite{Tan2022mbt}.

We should point out that similar Fermi surfaces or band structures were also reported in recent calculations in the MnBi$_2$Te$_4$-family materials \cite{PhysRevLett.122.206401,hu2020van,PhysRevLett.126.176403,PhysRevB.101.161109}. However, the  $\mathcal{C}_{6v}$-breaking was rarely appreciated and discussed with respect to the interaction with surface magnetism. 


\section{Conclusions \label{summary}}

The momentum-inversion symmetry-breaking provides us a criterion to identify the surface magnetism besides the surface magnetic gap.
This symmetry criterion may be more accessible than the magnetic gap, especially when the long-sought magnetic gap was not observed in ARPES results. 
To this end, we suggest that experiments should focus on the high-resolution energy dispersion along the $-K$ to $\Gamma$ to $K$ line and probe the symmetry-breaking without symmetrizing the ARPES data. Even in the presence of required surface magnetism, apparent symmetry-breaking exists only in selected energy windows where surface states are mainly contributed by the MnBi$_2$Te$_4$ layer.

\section{Acknowledgements}
We thank Shuolong Yang for helpful discussions. H.T. thanks the support from the Dean of Faculty Fellowship at Weizmann Institute of Science. B.Y. acknowledges the financial support from the European Research Council (ERC Consolidator Grant, No. 815869) and the Israel Science Foundation (ISF No. 3520/20).

%

\end{document}